\documentclass[aip,amssymb,reprint,superscriptadress,onecolumn]{revtex4-1}
\usepackage{graphicx}
\usepackage{lipsum}
\usepackage{hyperref}
\usepackage{amsmath}

\begin{document}

\title{Detection of DC currents and resistance measurements in longitudinal spin Seebeck effect experiments on Pt/YIG and Pt/NFO}

\author{Daniel Meier}
\email{dmeier@physik.uni-bielefeld.de}
\homepage{www.spinelectronics.de}
\author{Timo Kuschel}
\affiliation{Center for Spinelectronic Materials and Devices, Department of Physics, Bielefeld University, Universit\"atsstra\ss{}e 25, 33615 Bielefeld, Germany}
\author{Sibylle Meyer}
\author{Sebastian T.B. Goennenwein}
\affiliation{Walther-Meissner-Institut, Bayerische Akademie der Wissenschaften, Walther-Meissner-Strasse 8, 85748 Garching, Germany}
\author{Liming Shen}
\author{Arunava Gupta}
\affiliation{Center for Materials for Information Technology, University of Alabama, Tuscaloosa, Alabama 35487, USA}
\author{Jan-Michael Schmalhorst}
\author{G\"unter Reiss}
\affiliation{Center for Spinelectronic Materials and Devices, Department of Physics, Bielefeld University, Universit\"atsstra\ss{}e 25, 33615 Bielefeld, Germany}

\date{\today}

\begin{abstract}

In this work we investigated thin films of the ferrimagnetic insulators Y\(_{3}\)Fe\(_{5}\)O\(_{12}\) and NiFe\(_{2}\)O\(_{4}\) capped with thin Pt layers in terms of the longitudinal spin Seebeck effect (LSSE). The electric response detected in the Pt layer under an out-of-plane temperature gradient can be interpreted as a pure spin current converted into a charge current via the inverse spin Hall effect. Typically, the transverse voltage is the quantity investigated in LSSE measurements (in the range of \(\mu V\)). Here, we present the directly detected DC current (in the range of nA) as an alternative quantity. Furthermore, we investigate the resistance of the Pt layer in the LSSE configuration. We found an influence of the test current on the resistance. The typical shape of the LSSE curve varies for increasing test currents.

\end{abstract}

\maketitle

In the recent years, the spin Seebeck effect (SSE), the thermal generation of a pure spin current, has been attracted much attention in spintronics\cite{Hoffmann:2015ab} and has opened the branch of spin caloritronics.\cite{Bauer:2012fq} The first observation on thin Ni\(_{81}\)Fe\(_{19}\) (permalloy - Py) films\cite{Uchida:2008cc} in the now called transverse configuration (TSSE) could not be reproduced by many groups.\cite{Huang:2011cd,Avery:2012bj,Schmid:2013m,Meier:2013fa} Additionally, unintended charge transport phenomena like the anomalous Nernst or planar Nernst effect appeared in most attempts to investigate the TSSE which disguised the voltage measured. These unintended Nernst effects are effected by temperature gradients in unintended directions\cite{Meier:2013fa} or small parasitic magnetic fields.\cite{Shestakov:2015ab}

Magnetic insulators like Y\(_3\)Fe\(_5\)O\(_{12}\) (yttrium iron garnet - YIG) or NiFe\(_2\)O\(_4\) (nickel ferrite - NFO) seem to be a more promising material class for TSSE investigations.\cite{Uchida:2010ei} The lack of free charge carriers suppress the appearance of any thermally driven charge current phenomena. However, it could be shown very recently that the TSSE on YIG and NFO could also not be reproduced.\cite{Meier:2015fe} Nevertheless, a pure spin current generation could be observed in spite of that. This was accomplished by a spin Seebeck effect in the longitudinal configuration (LSSE).\cite{Uchida:2010jb} Here, the spin current is generated longitudinal to the temperature gradient which is applied perpendicular to the spin detector/ferromagnet bilayer system. 

The LSSE on ferromagnetic or ferrimagnetic insulators is now a well established phenomena and could be reproduced in many groups.\cite{Huang:2012tk,Qu:2013jj,Meier:2013dz,Kehlberger:2014ha,Schreier:2014cc} The generally used quantity which is presented in all of the given publications is the voltage which arises in the spin detector material transverse to the generated spin current due to the inverse spin Hall effect (ISHE). The ISHE describes the conversion of a spin current into a charge current due to spin dependent scattering of the electrons in a heavy metal.\cite{Saitoh:2006kk} Generally, the voltage is used as the electrical response quantity that is measured in ISHE experiments. Recently, Omori et al. have investigated the detection of the converted charge current in lateral spin valve structures by means of the ISHE.\cite{Omori:2014hp} The direct detection of the charge current generated in the spin detector material in LSSE experiments will be presented in this work.

Another transport phenomena which is connected with Pt/magnetic insulator bilayers is the recently observed spin Hall magnetoresistance (SMR).\cite{Nakayama:2013gs,Althammer:2013ck} Here, an interplay of the spin Hall effect and the ISHE leads to a magnetoresistance effect when an electrical current flows through the Pt film. Most of the given literature show field rotation measurements to distinguish between the SMR and the anisotropic magnetoresistance. The latter effect could appear if the Pt is spin polarized at the interface by the ferromagnetic or ferrimagnetic material due to a magnetic proximity effect.\cite{Huang:2012tk} This is still under discussion for the investigated Pt/YIG systems\cite{Lu:2013eq,Geprags:2012be}, but could be excluded for Pt/NFO by x-ray resonant magnetic reflectivity\cite{Kuschel:2015hv,Kuschel:2015ab} and for Pt/CoFe\(_2\)O\(_4\) by x-ray magnetic circular dichroism very recently.\cite{Valvidares:2015vu} The SMR can also emerge as symmetric peaks in magnetic field loop measurements due to magnetic anisotropies. 

\begin{figure}[!h]
\centering
\includegraphics[width=\linewidth]{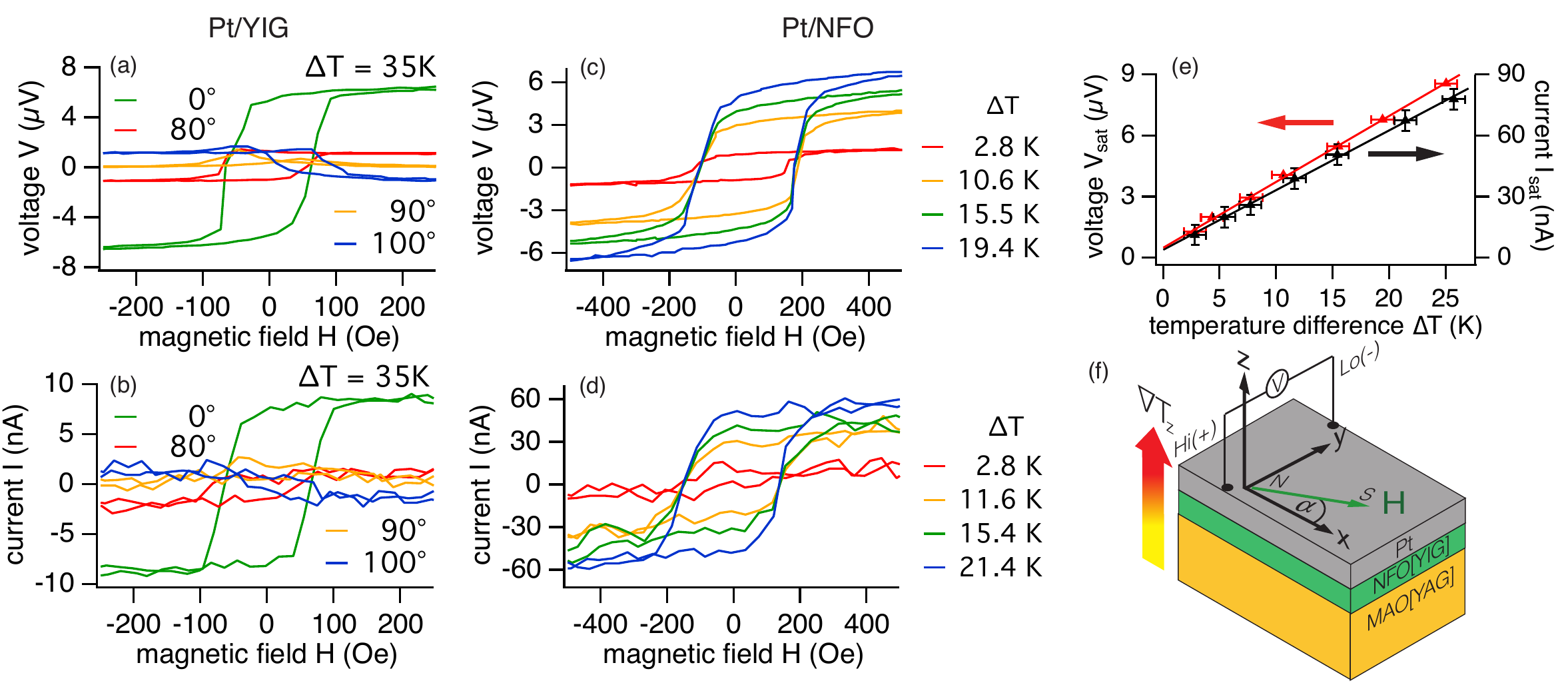}
\caption{(a) The transverse voltage \(V\) is shown as a function of the external magnetic field \(H\) for various angles \(\alpha\) with respect to the x-direction measured on Pt/YIG. The temperature difference between the top and bottom is \(\Delta T = 35\,K\). (b) The transverse current \(I\) measured at the Pt/YIG bilayer is shown as a function of \(H\) for various angles \(\alpha\). (c) The voltage \(V\) shown as a function of \(H\) for various temperature differences \(\Delta T\) on Pt/NFO. (d) The current \(I\) measured at the Pt/NFO bilayer plotted against \(H\). (e) The saturated voltage \(V_{sat}\) and  saturated current \(I_{sat}\) shown as a function of \(\Delta T\) for Pt/YIG. (f) The measurement configuration with the temperature gradient \(\nabla T_z\), the magnetic field vector \(\textbf{H}\) and the connections as well as polarity of the electrical measurement.}
\label{figure-1}
\end{figure}

In this work we present measurements on Pt/YIG and Pt/NFO in the LSSE configuration (Fig.\,\ref{figure-1}\,(f)). The electric response detected in the Pt layer is shown as a function of the external magnetic field \(H\) for various angles \(\alpha\) with respect to the x-direction. Here, we present the voltage as the typically used quantity for reference and compare this with the directly detected DC current as an alternative quantity. Furthermore, we show how the resistance varies during the LSSE measurement. We could observe different behaviour when the test current is increased. Large test currents lead to a dominant SMR which suppresses the appearance of the LSSE. Therefore, we will show that the LSSE and the SMR can be created simultaneously with different sources. While the LSSE is generated by the temperature gradient, the SMR is produced by a charge current through the Pt. Furthermore, we used the same contacts for the applied current and the voltage measurement. Slightly different experiments are reported by Schreier et al.\cite{Schreier:2013kl} and Vlietstra et al.\cite{Vlietstra:2014es} about simultaneous LSSE/SMR measurements with the same source.

The \(t_{YIG}\)\,=\,60\,nm thick YIG film investigated in this work was deposited on 0.5\,mm thick yttrium aluminium garnet (Y\(_3\)Al\(_5\)O\(_{12}\)) (111)-oriented single crystal substrates with 5\,mm\,\(\times\)\,2\,mm in dimension by pulsed laser deposition from a stoichiometric polycrystalline target.\cite{Althammer:2013ck} The KrF excimer laser had a wavelength of 248\,nm, a repetition rate of 10\,Hz and an energy density of 2\,J/cm\(^2\). The YIG film was capped by a 2\,nm thin Pt film deposited by e-beam evaporation. The \(t_{NFO}\)\,=\,1\,\(\mu\)m thick NFO film was deposited by direct liquid injection-chemical vapour deposition on 0.5\,mm thick MgAl\(_2\)O\(_4\) (100)-oriented substrates with 8\,mm\,\(\times\)\,5\,mm in dimension.\cite{Li:2011ka} The NFO film was cleaned with ethanol in an ultrasonic bath after a vacuum break and was capped by a 10\,nm thin Pt film deposited by dc magnetron sputtering.

The LSSE measurements were performed in a vacuum chamber with a base pressure of \(1 \cdot 10^{-6}\)\,mbar. The samples were clamped between two copper blocks with a piece of 0.5\,mm thick sapphire substrate for electrical isolation between the Pt and the top copper block. The temperature gradient through the sample stack was established by heating the top copper block by Joule heating. Two \(25\,\mu\)m thin aluminium bonding wires at the sample edges measured the electrical response transverse to the temperature gradient and perpendicularly to the external magnetic field which was applied in the sample plane. For Pt/YIG and Pt/NFO the electrical contacts had a distance of about 4\,mm and 7\,mm, respectively. The measurement configuration and all observed responses are consistent if a rigorous sign check is applied.\cite{Schreier:2014cc}

In the first LSSE measurements shown in Fig.\,\ref{figure-1}\,(a) the voltage \(V\) on the Pt/YIG bilayer was obtained as a function of the external magnetic field \(H\) with a fixed temperature difference \(\Delta T = 35\,K\) for various angles \(\alpha\) of the magnetic field vector with respect to the x-direction. The voltage in saturation is about \(6.8\mu V\) for \(\alpha = 0^{\circ}\) which decreases for angles up to \(\alpha = 90^{\circ}\) where it reaches zero due to the cross product of the ISHE given by \({\bf{E}} \propto {\bf{J}}_S \times \boldsymbol{\sigma}\), with the electric field {\bf{E}}, the spin current \({\bf{J}_S}\) and the spin-polarization vector \({\boldsymbol{\sigma}}\) of the electrons in the Pt. For angles \(\alpha\) above \(90^{\circ}\) the voltage in saturation changes its sign. For magnetic field values \(H\) around the coercive field of the YIG the voltage also switches its sign. This can be seen for all angles \(\alpha\). However, for angles around \(90^{\circ}\) there are two peaks around the coercive field. These peaks results from a switching behaviour for materials with a magnetic anisotropy which was investigated recently.\cite{Kehlberger:2014ha} While Kehlberger et al. could observe an antisymmetric switching with respect to H we observe a symmetric switching. This can be a result of a symmetric reversal process of the magnetization vector which rotates between \(0^\circ\) and \(180^\circ\) passing the \(90^\circ\) direction for both hysteresis branches and never rotating over the \(270^\circ\) direction. Additionally, this symmetric curve can be reminiscent to a magnetoresistive switching in a manner like the SMR and will be part of future investigations. In Fig.\,\ref{figure-1}\,(b) the current \(I\) measured at the Pt contacts is plotted as a function of \(H\). The current generated by the ISHE conversion in the Pt shows the same angle dependency of the saturated value compared to the voltage. The current in saturation decreases for angles \(\alpha\) between \(0^\circ\) and \(90^\circ\) until it vanishes. This shows the same behaviour expected for the LSSE via the ISHE. The peaks in the voltage for angles \(\alpha\) near \(90^\circ\) (Fig.\,\ref{figure-1}\,(a)) are vaguely perceptible in the current due to the different measurement accuracy.

In the additional system Pt/NFO the current \(I\) and the voltage \(V\) generated by the LSSE were studied. A detailed LSSE investigation for this system was previously reported.\cite{Meier:2013dz} In Fig.\,\ref{figure-1}\,(c) \(V\) is shown as a function of \(H\) for various temperature differences \(\Delta T\). The magnitude of these curves is shown in Fig.\,\ref{figure-1}\,(e) which shows the typical proportionality expected for the LSSE. This could be confirmed for the current \(I\) directly measured at the Pt film (Fig.\,\ref{figure-1}\,(d)) for slightly divergent temperature differences. However, the linearity between \(I\) and \(\Delta T\) becomes obvious in Fig.\,\ref{figure-1}\,(e) in spite of the poorer accuracy. Therefore, the current \(I\) shows the expected LSSE behaviour for the \(\Delta T\) proportionality as well as the angle dependency regarding the ISHE exemplarily shown for each material system, Pt/YIG and Pt/NFO. This makes it an equivalent quantity for LSSE investigations.

\begin{figure}[!h]
\centering
\includegraphics[width=\linewidth]{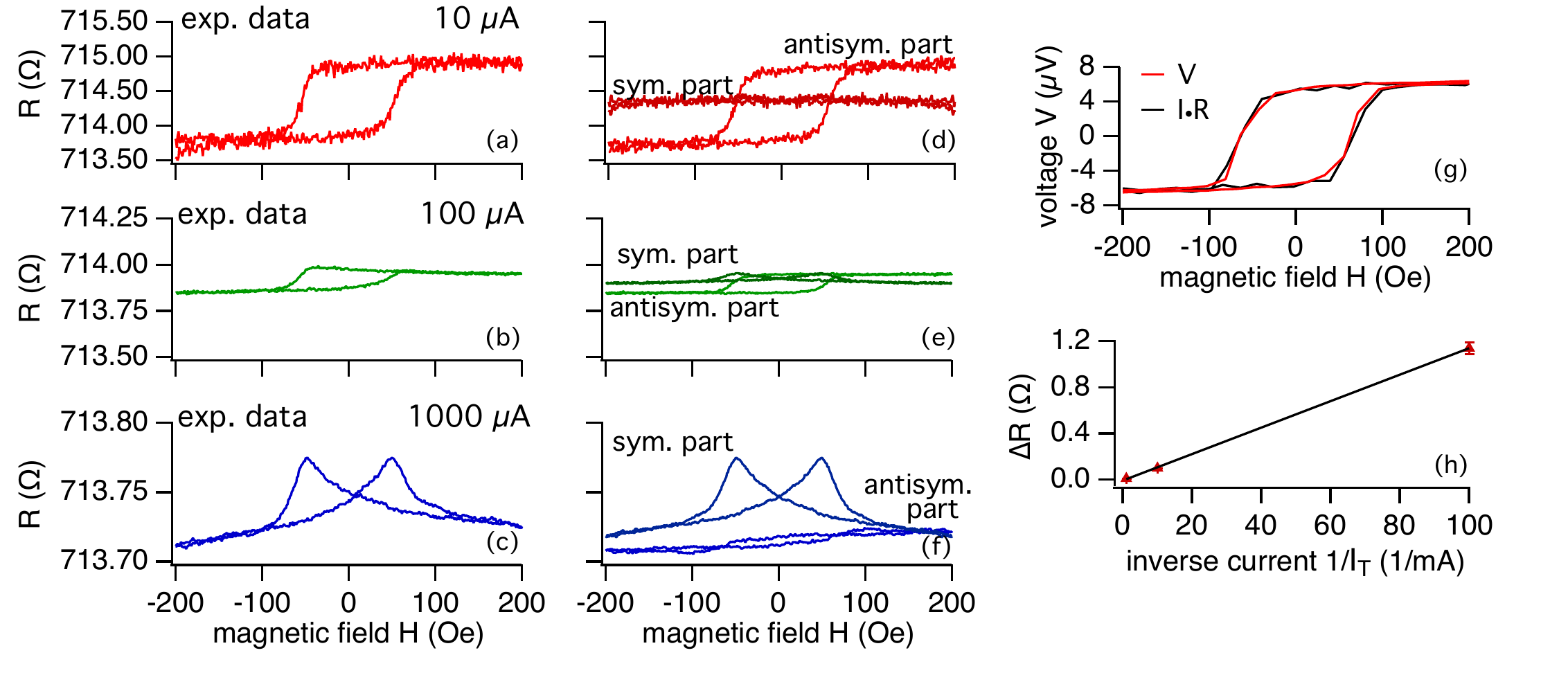}
\caption{The resistance \(R\) measured on Pt/YIG transverse to the applied external magnetic field \(H\) for \(\Delta T = 35\,K\) with different test currents \(I_T\) in (a) 10\,\(\mu\)A, (b) 100\,\(\mu\)A and (c) 1000\,\(\mu\)A. The experimental data was separated mathematically into the antisymmetric and symmetric part (d), (e) and (f) with respect to \(H\). (g) The directly measured LSSE voltage compared to the product of the measured DC current and the residual resistance of the Pt film. (h) The margin of the saturated values in the antisymmetric parts \(\Delta R\) plotted against the inverse test current \(1/I_T\).}
\label{figure-2}
\end{figure}

In further investigations, we measured the resistance of the bilayers for \(\Delta T\)\,=\,35\,K. The measurement is performed as a voltage measurement with different test currents applied. In Fig.\,\ref{figure-2}\,(a), (b) and (c) the resistance \(R\) is shown as a function of \(H\) for three different test currents (10\,\(\mu\)A, 100\,\(\mu\)A and 1000\,\(\mu\)A). For low test currents, e.g., 10\,\(\mu\)A the resistance is antisymmetric with respect to \(H\) and shows a hysteretical behaviour. The experimental data can be separated mathematically into a complete antisymmetric and symmetric part which is shown in Figs.\,\ref{figure-2}\,(d), (e) and (f). Since the experimental data are nearly completely antisymmetric for \(I_T\)\,=\,10\,\(\mu\)A the symmetric part shows only a mean resistance without any magnetic field dependent behaviour (Fig.\,\ref{figure-2}\,(d)). For larger test currents, however, the mathematical separation shows a switching behaviour with two peaks around the coercive fields of the YIG film which is symmetric with \(H\) (Fig.\,\ref{figure-2}\,(e)). The magnitude of the antisymmetric part, i.e., the difference of the saturated voltages for positive and negative magnetic fields \(\Delta R\), decreases. When the test current is further increased the symmetric effect is more dominant in the experimental data compared to the antisymmetric contribution. Here, the antisymmetric part shows a very low magnitude (Fig.\,\ref{figure-2}\,(f)) and the symmetric part is more dominant even in the experimental data. The LSSE contribution normalized to the used test current is always the same. This can be shown by the proportionality between \(\Delta R\) and the inverse test current 1/\(I_T\) in Fig.\,\ref{figure-2}\,(h). When the the measured ISHE current (Fig.\,\ref{figure-1}\,(a)) is multiplied by the residual resistance of the Pt film the obtained curve is similar to the previously measured ISHE voltage (Fig.\,\ref{figure-2}\,(g)).

Recently, Schreier et al. have shown that the temperature gradient can also be established by heating the top of the sample with a large current through the Pt layer which is the spin detector at the same time.\cite{Schreier:2013kl} The Joule heating of a large d.c. current (about 10\,mA) transverse to the voltage measurement generated the LSSE which is antisymmetric with the external magnetic field \(H\). Furthermore, the current generated a SMR represented by large symmetric peaks. Both could be separated by taking the difference of two measurements with the reversed current applied at the Pt. For the test currents used in this work there is no additional heating which would be manifest in a deviation of the linearity between \(\Delta R\) and \(1/I_T\) in Fig.\,\ref{figure-2}\,(h). 

Very recently, Vlietstra et al. have extended these investigations by using a.c. currents in a similar range of the absolute value compared to Schreier et al. They measured the first- and second-harmonic voltage in order to separate the SMR and LSSE contribution which are generated by the same current source.\cite{Vlietstra:2014es}

In conclusion, we have shown that the direct measurement of the d.c. current is an equivalent quantity in LSSE experiments which shows the same properties of the saturated values compared to the commonly used voltage. Furthermore, the resistance was measured in the LSSE configuration by applying different test currents. The same switching behaviour expected for the LSSE could be obtained for low enough test currents applied. However, large enough test currents can affect the result and create an additional contribution given by the SMR. The variation of the test current can obscure the real interpretation of magnetoresistive experiments which become extremely worthwhile for the sample systems used in LSSE experiments.

The authors thank Michael Schreier for valuable discussions and gratefully acknowledge financial support by the EMRP JRP EXL04 SpinCal and the Deutsche Forschungsgemeinschaft (DFG) within the priority programme SpinCaT (KU 3271/1-1 and RE 1052/24-2).

\bibliographystyle{apsrev4-1}
%

\end{document}